\documentclass[aps,prl,twocolumn,showpacs,superscriptaddress,floatfix]{revtex4-1}
\usepackage{amsmath,amssymb}
\usepackage{graphicx}
\usepackage{epsfig}
\usepackage{color}
\usepackage{rotating}
\usepackage{bm}
\usepackage{setspace}
\usepackage{hyperref}
\begin{document}
%Title of paper
\title{Imaging the Electron-Boson Coupling in Superconducting FeSe}
\author{Can-Li Song}
\affiliation{State Key Laboratory for Surface Physics, Institute of Physics, Chinese Academy of Sciences, Beijing 100190, China}\affiliation{State Key Laboratory of Low-Dimensional Quantum Physics, Department of Physics, Tsinghua University, Beijing 100084, China}\affiliation{Department of Physics, Harvard University, Cambridge, Massachusetts 02138, U. S. A}
\author{Yi-Lin Wang}
\affiliation{State Key Laboratory for Surface Physics, Institute of Physics, Chinese Academy of Sciences, Beijing 100190, China}
\author{Ye-Ping Jiang}
\affiliation{State Key Laboratory for Surface Physics, Institute of Physics, Chinese Academy of Sciences, Beijing 100190, China}\affiliation{State Key Laboratory of Low-Dimensional Quantum Physics, Department of Physics, Tsinghua University, Beijing 100084, China}
\author{Zhi Li}
\author{Lili Wang}
\author{Ke He}\affiliation{State Key Laboratory for Surface Physics, Institute of Physics, Chinese Academy of Sciences, Beijing 100190, China}
\author{Xi Chen}
\affiliation{State Key Laboratory of Low-Dimensional Quantum Physics, Department of Physics, Tsinghua University, Beijing 100084, China}
\author{Jennifer E. Hoffman}
\affiliation{Department of Physics, Harvard University, Cambridge, Massachusetts 02138, U. S. A}
\author{Xu-Cun Ma}
\email[]{xcma@iphy.ac.cn}
\affiliation{State Key Laboratory for Surface Physics, Institute of Physics, Chinese Academy of Sciences, Beijing 100190, China}
\author{Qi-Kun Xue}
\email[]{qkxue@mail.tsinghua.edu.cn}
\affiliation{State Key Laboratory for Surface Physics, Institute of Physics, Chinese Academy of Sciences, Beijing 100190, China}\affiliation{State Key Laboratory of Low-Dimensional Quantum Physics, Department of Physics, Tsinghua University, Beijing 100084, China}
\date{\today}

\begin{abstract}
Scanning tunneling spectroscopy has been used to reveal signatures of a bosonic mode in the local quasiparticle density of states of superconducting FeSe films. The mode appears below $T_c$ as a `dip-hump' feature at energy $\Omega\sim 4.7 k_B T_c$ beyond the superconducting gap $\Delta$. Spectra on strained regions of the FeSe films reveal simultaneous decreases in $\Delta$ and $\Omega$. This contrasts with all previous reports on other high-$T_c$ superconductors, where $\Delta$ locally anti-correlates with $\Omega$. A local strong coupling model is found to reconcile the discrepancy well, and to provide a unified picture of the electron-boson coupling in unconventional superconductors.
\end{abstract}
%\maketitle must follow title, authors, abstract, \pacs, and \keywords
\maketitle

Iron-based superconductors (Fe-SCs), due to their relatively high transition temperature $T_{c}$ and resemblance to the even higher-$T_{c}$ cuprates, have evoked tremendous excitement and renewed hope for unveiling the microscopic pairing mechanism of high-$T_{c}$ superconductivity \cite{kamihara2008iron}. Similar to cuprate and heavy fermion materials, the superconductivity in most Fe-SCs emerges in close proximity to an antiferromagnetic order, suggesting the relevance of spin fluctuations to the pairing glue \cite{Mazin2008, Kuroki2008}. In support of this hypothesis, inelastic neutron scattering experiments have revealed a spin resonance at the nesting wavevector connecting the $\Gamma$-centered hole pockets to the $M$-centered electron pockets in iron pnictides \cite{de2008magnetic,lumsden2010magnetism,taylor2011antiferromagnetic} as well as FeSe$_{0.4}$Te$_{0.6}$ \cite{qiu2009spin}. However, the recently-discovered superconductivity in ternary iron selenides $A$Fe$_2$Se$_2$ ($A$=K, Cs, Rb, or Tl) without $\Gamma$-centered hole pockets \cite{Qian2011Absence, Mou2011Distinct, zhang2011nodeless}, together with the unexpected robustness of $T_{c}$ against impurities in Fe-SCs \cite{Kawabata2008super}, challenges this picture and returns attentions to other pairing candidates, such as phonon and phonon/magnetism-induced orbital fluctuations \cite{kontani2010orbital, Shimojima2011orbital, Chu2013iron}. The superconducting mechanism in Fe-SCs thus remains enigmatic.

Tunneling experiments can provide crucial insights into the pairing mechanism. The mediator (or bosonic mode), which binds electrons into superconducting Cooper pairs, interacts with electrons and thus reconstructs the quasiparticle density of states (DOS). The DOS reconstruction can be probed by the differential tunneling conductance $dI/dV$, in which the features of the bosonic mode often appear at energies $\pm(\Delta+\Omega)$, where $\Delta$ is the superconducting gap energy and $\Omega$ is the energy of the bosonic mode. Such experiments have unequivocally established the electron-phonon mechanism of conventional superconductivity in materials such as Pb \cite{mcmillan1965lead}. For cuprates and the more recently discovered iron pnictides, measurements with scanning tunneling microscopy and spectroscopy (STM/STS) revealed dip-hump structures at energies above $\Delta$, which have been controversially interpreted as fingerprints of phonon or spin fluctuations \cite{zasadzinski2001correlation, lee2006interplay, niestemski2007distinct, jenkins2009imaging, fasano2010local, shan2012evidence, wang2012close, chi2012scanning}. Furthermore, such STM/STS studies in cuprates and iron-pnictides consistently revealed a surprising anti-correlation between $\Delta$ and $\Omega$ \cite{niestemski2007distinct, jenkins2009imaging, fasano2010local, shan2012evidence, lee2006interplay}. This has been tentatively accounted for by a local strong coupling model \cite{balatsky2006local}, although the key parameter ranges of this model have not been experimentally verified.

On the other hand, no such collective mode has yet been experimentally elucidated in the structurally simplest 11-type binary superconductor FeSe  \cite{hsu2008superconductivity}. The lack of large stoichiometric FeSe single crystals poses tremendous barriers to obtaining such information by inelastic neutron scattering, although previous studies have raised interesting questions. These include, for example, a different magnetic order for the parent compound FeTe compared to other FeAs-based parent compounds \cite{Li2009First}, the occurrence of superconductivity in stoichiometric FeSe without doping, and reports of $T_{c}$ up to 65 K in single-unit-cell FeSe films on SiTiO$_3$ which completely lack  $\Gamma$-centered Fermi pockets \cite{wang2012interface, liu2012electronic, tan2013interface}. Moreover, $dI/dV$ spectra on multilayer FeSe films on graphitized SiC(0001) revealed evidence for a gap function with nodal lines \cite{song2011direct}, in stark contrast to the nodeless superconducting gap in single unit cell FeSe films on SiTiO$_3$ \cite{wang2012interface, liu2012electronic} and in closely related FeSe$_{0.4}$Te$_{0.6}$ \cite{hanaguri2010unconventional}. Thus considerable concerns over the interplay between electronic structure, phonons, magnetism, and superconductivity have emerged.

Here we report the STM observation of a bosonic mode occurring outside the superconducting gap, and its coupling with electrons in stoichiometric and superconducting FeSe films. All experiments are conducted on a Unisoku ultrahigh vacuum STM system equipped with molecular beam epitaxy (MBE) for film growth. High-quality FeSe films are obtained by co-evaporating high-purity Fe (99.995$\%$) and Se (99.999$\%$) onto graphitized 6\textit{H}-SiC(0001) substrate, as detailed elsewhere \cite{song2011direct, song2011molecular, song2012suppression}. A polycrystalline PtIr tip is used throughout the experiments. Tunneling spectra are measured by disabling the feedback circuit at setpoint voltage $V_{\mathrm{s}}$ = 10 mV and $I$ = 0.2 nA, sweeping the sample voltage $V$, and extracting the differential conductance $dI/dV$ using a standard lock-in technique with a small bias modulation of 0.1 mV at 987.5 Hz.

\begin{figure}[tb]
\includegraphics[width=\columnwidth]{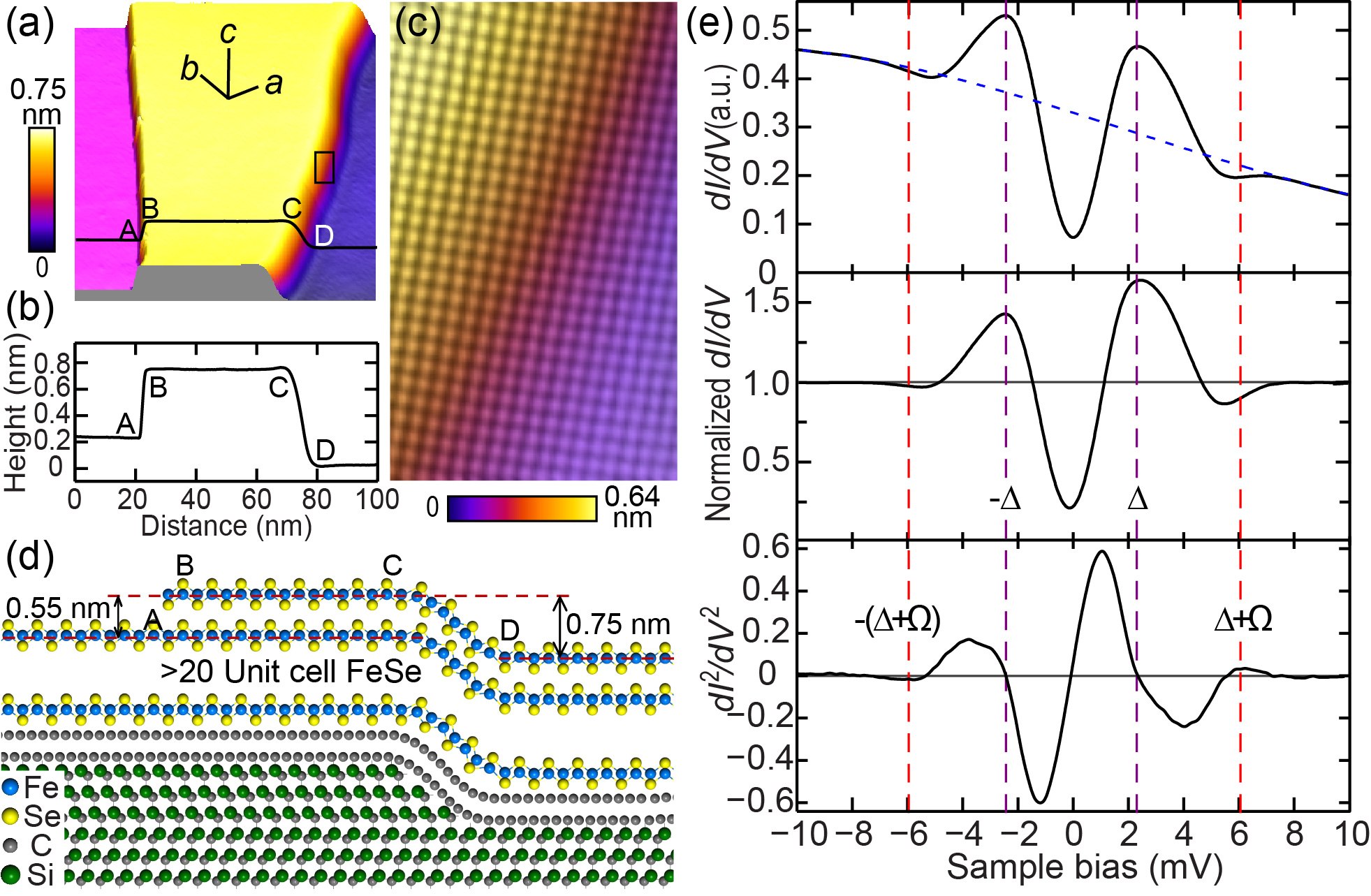}
\caption{(a) Large-scale STM topographic image of a FeSe film ($V_{\mathrm{s}}$ = 2.5 V, $I$ = 0.1 nA, 100 nm $\times$ 100 nm). (b) Profile taken along the black curve in (a). (c) Atomically resolved STM image on the region marked by black rectangle in (a) ($V_{\mathrm{s}}$ = 1.0 mV, $I$ = 0.1 nA, 6 nm $\times$ 10 nm). (d) Cross-sectional schematic representation of the formation of the two different steps in (a). The 6\textit{H}-SiC(0001) unit cell consists of six bilayers (ABCACB stacking) with sequences of three linearly stacked bilayers followed by an orientation change. (e) Black traces show raw $dI/dV$ (top panel), normalized $dI/dV$ (middle panel) and $d^2I/dV^2$ (bottom panel) spectra on the flat terraces at 3.0 K. The normalization was performed by dividing the raw $dI/dV$ spectrum by its background, which was extracted from a cubic fit to the conductance for $|V | > 8$ mV (blue dashes in the top panel). The purple and red dashes show the approximate energy positions of $\pm\Delta$, $\pm(\Delta+\Omega)$, respectively.
}
\end{figure}

Figure 1(a) depicts a constant-current topographic image of the as-grown FeSe(001) films. Three color-coded terraces (from left to right: magenta, yellow, and dark blue) are divided by two steps `AB' and `CD' with different heights, as clearly revealed by the line profile in Fig.\ 1(b). The left step AB has a height of 0.55 nm, equal to the $c$-axis lattice constant (0.5518 nm) of FeSe \cite{hsu2008superconductivity}, while the right step CD is higher, 0.75 nm. Shown in Fig.\ 1(c) is a zoom-in STM image of the CD step region, marked by the black rectangle in Fig.\ 1(a). Intriguingly, the atomic lattice is continuous, indicative of a physically continuous FeSe films across the CD step. Additionally, the step height of 0.75 nm matches with three SiC bilayers (0.25 nm each). We thus propose that the two distinct steps are formed by the atomic configuration sketched in Fig.\ 1(d). Here the CD step stems from the SiC substrate, and FeSe films can continuously straddle the underlying SiC step. This will certainly lead to strain and may alter the superconductivity in the step region.

Figure 1(e) typifies the $dI/dV$ tunneling spectrum measured on the flat terraces. In addition to the superconducting gap, dip-hump features are visible outside the coherence peaks, with $E_{\mathrm{dip}}\thicksim\pm$5.7 meV (top panel). They are more clearly seen in the normalized $dI/dV$ spectrum (middle panel), and bear striking resemblance to those observed in cuprates and iron pnictides \cite{zasadzinski2001correlation, niestemski2007distinct, jenkins2009imaging, fasano2010local, shan2012evidence, wang2012close, chi2012scanning, lee2006interplay}. In analogy to these previous studies, we assign the dip-hump features to the coupling of the quasiparticles with a collective bosonic excitation. To read out the mode energy, we calculate numerically the second derivative of conductance $d^2I/dV^2$, shown in the bottom panel of Fig.\ 1(e). We take the maximum at positive voltage and minimum at negative voltage, marked by the red dashes, as excellent estimates of the energies $\pm(\Delta + \Omega$) ($\sim\pm6.0$ meV) \footnote{Similar work in cuprate superconductors has also used the extrema in $d^2I/dV^2$ as approximate markers of $\pm(\Delta+\Omega)$ \cite{niestemski2007distinct, lee2006interplay}, however, an exact assignment would require an integral of the unknown boson band structure over the full Brillouin zone \cite{mcmillan1965lead, song2013pairing}.}. $\Delta$ is measured as half the peak-to-peak energy separation in $dI/dV$ ($\sim2.2$ meV). We thereby extract the bosonic mode energy $\Omega\sim 3.8\pm0.1$ meV.

\begin{figure}[tb]
 \includegraphics[width=\columnwidth]{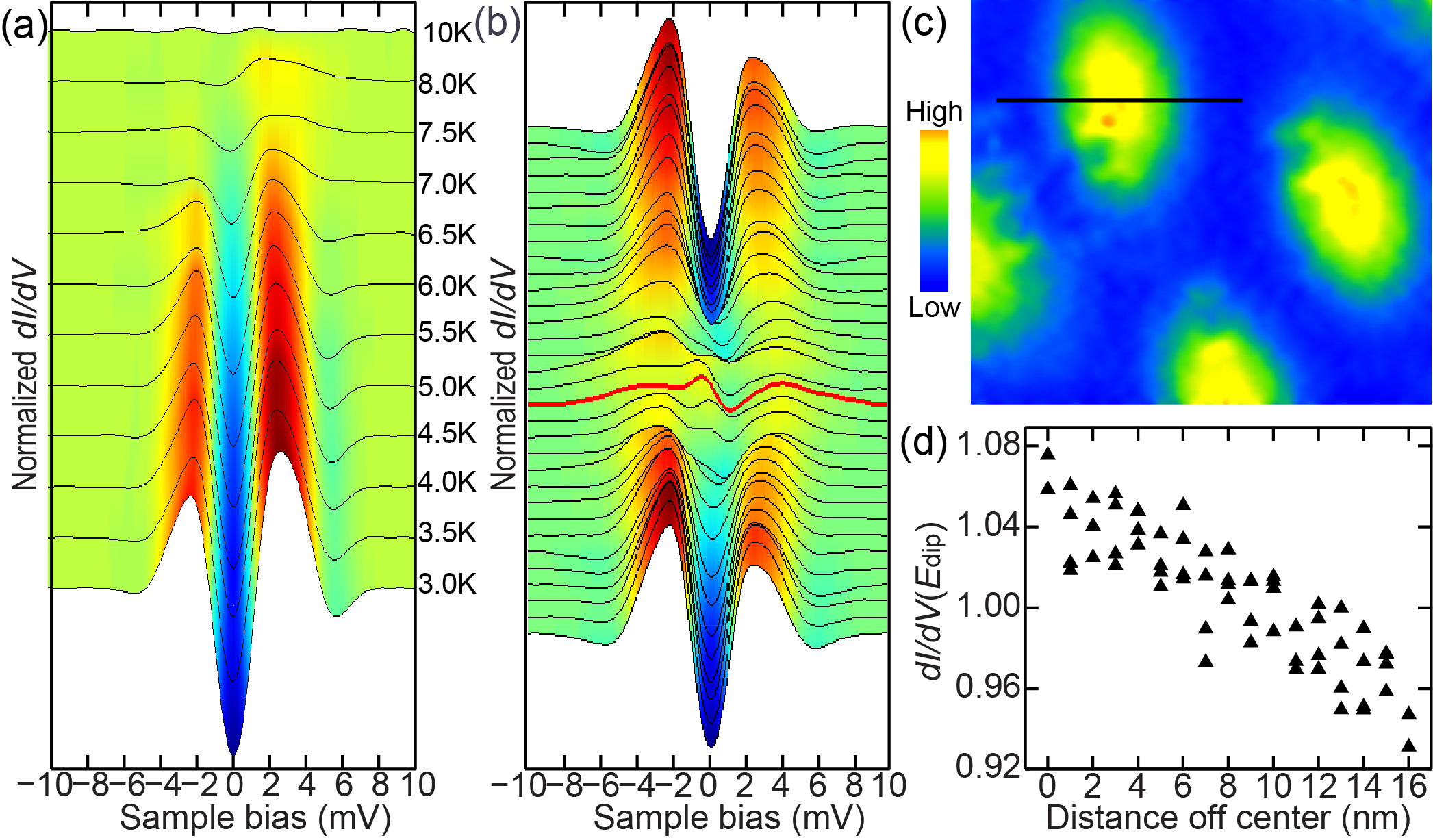}
  \caption{Temperature and magnetic field dependence of conductance spectra away from step edges. (a) Normalized $dI/dV$ spectra at temperatures ranging from 3.0 K to 10 K ($T_{c}\sim9.3$ K). (b) A series of normalized $dI/dV$ spectra straddling a single magnetic vortex in (c). The spectra are equally spaced 1.0 nm apart along a 32 nm trajectory, and measured at 4.5 K. The thick red line emphasizes the spectrum at the vortex center. (c) A 60 nm $\times$ 52 nm zero bias conductance (ZBC) map showing four vortices at 4.5 K and 2 T. The black line indicates the trajectory along which the spectra shown in (b) are measured. (d) Site dependence of $dI/dV$ at energy $E_{\mathrm{dip}}=\pm$5.7 meV from (c).
  }
\end{figure}

To determine whether this bosonic mode links with the superconducting state in FeSe, we explored the temperature and magnetic field dependence of the tunneling spectra. Figure 2(a) presents the evolution of the normalized $dI/dV$ spectra with temperature up to 10 K, just above $T_{c} \sim 9.3$ K \cite{song2011molecular}. As can be seen, the bosonic excitations and superconducting gap progressively vanish near $T_{c}$. This provides the first evidence that the bosonic mode is closely related with the superconductivity. Furthermore, one can investigate the interplay between the bosonic excitations and superconductivity around magnetic vortices. When the magnetic field enters the superconducting FeSe films in the form of vortices, it destroys the superconductivity inside the vortex cores \cite{song2011direct, song2012suppression}.  Figure 2(b) shows a series of normalized $dI/dV$ spectra cutting through one vortex in Fig.\ 2(c). The vortices are imaged by mapping zero bias conductance (ZBC) at 2 T magnetic field applied perpendicular to the sample surface. Figure 2(d) plots the site dependence of normalized $dI/dV$ at $E_{\mathrm{dip}}=\pm$5.7 meV. Smaller $dI/dV$($E_{\mathrm{dip}}$) means a stronger electron-boson coupling \cite{pasupathy2008electronic}. Apparently, both the bosonic excitations and superconducting gap gradually broaden out on approaching the vortex center. All evidence consistently indicates an intimate connection between the observed bosonic mode and superconductivity, suggesting that this mode may act to `glue' electrons together to form Cooper pairs in FeSe.

\begin{figure}[tb]
  \includegraphics[width=0.835\columnwidth]{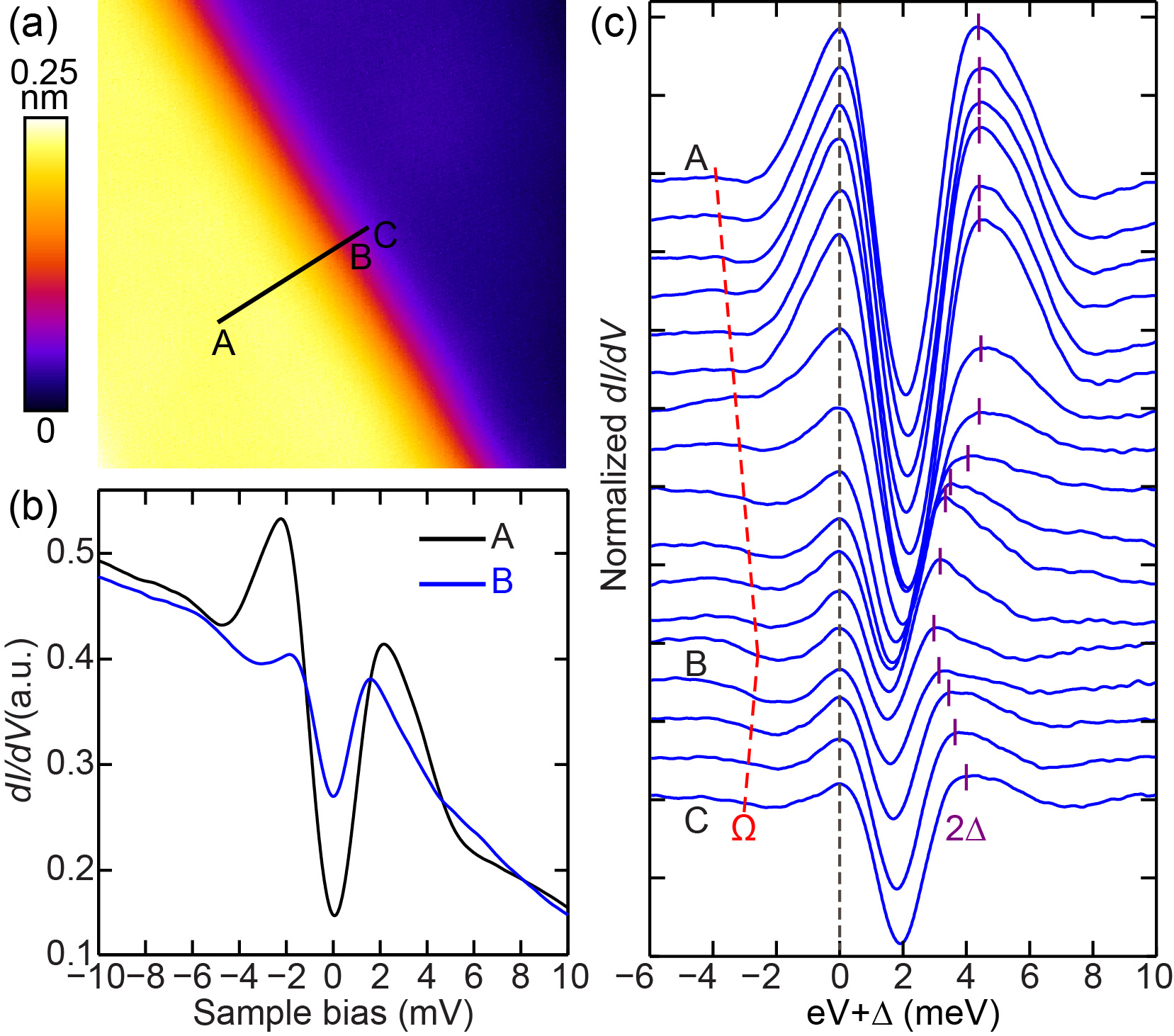}
  \caption{(a) STM topography of FeSe film across a single SiC bilayer step ($V_{\mathrm{s}}$ = 2.0 V, $I$ = 0.2 nA, 45 nm $\times$ 45 nm). (b) Raw $dI/dV$ spectra for site A (black) without strain and site B (blue) with strain in (a). (c) Normalized $dI/dV$ spectra acquired 1 nm apart along the line ABC in (a). All spectra were measured at 4.5 K. Each spectrum has been shifted to the right by an energy corresponding to its gap magnitude $\Delta$, in order to visually separate the spatial dependence of $\Omega$ (red) from $2\Delta$ (purple). The dashed gray line is a guide to the eye.
  }
\end{figure}

To bring more insight into the observed bosonic mode, we quantify its relationship to the superconducting gap $\Delta$, by studying the tunneling spectra on the strained regions. Figure 3(a) shows a topographic image of another continuous FeSe film straddling a single SiC bilayer. Such films bend downwards in the step region, resulting in tensile strain at the top surface, which enables a local measurement of the strain-tailored superconducting properties and bosonic excitations. Figure 3(b) displays the $dI/dV$ spectra in regions with and without strain. In the strained region, $\Delta$ decreases, ZBC is elevated and the superconducting coherence peaks are strongly suppressed, compared to the strain-free region. These observations demonstrate the suppression of superconductivity by tensile strain, consistent with previous transport measurements \cite{nie2009suppression}. Furthermore, the dip-hump features shift toward the Fermi level, indicating the reduction of $\Delta + \Omega$. To reveal the relationship between $\Omega$ and $\Delta$ clearly, Fig.\ 3(c) plots the normalized $dI/dV$ spectra across the step, with the energy shifted by $\Delta$. The red dashed lines indicate the energies of $\Omega$, while the short purple lines the energies of $2\Delta$. Evidently, $\Omega$ increases with increasing $\Delta$. Figure 4(a) summarizes the gap magnitude $\Delta$ versus $\Omega$. Despite some scatter in the data, it is clear that $\Delta$ correlates positively with $\Omega$. This supports the contribution of the observed bosonic mode to electron pairing and superconductivity in FeSe.

\begin{figure*}[tb]
 \includegraphics[width=1.5\columnwidth]{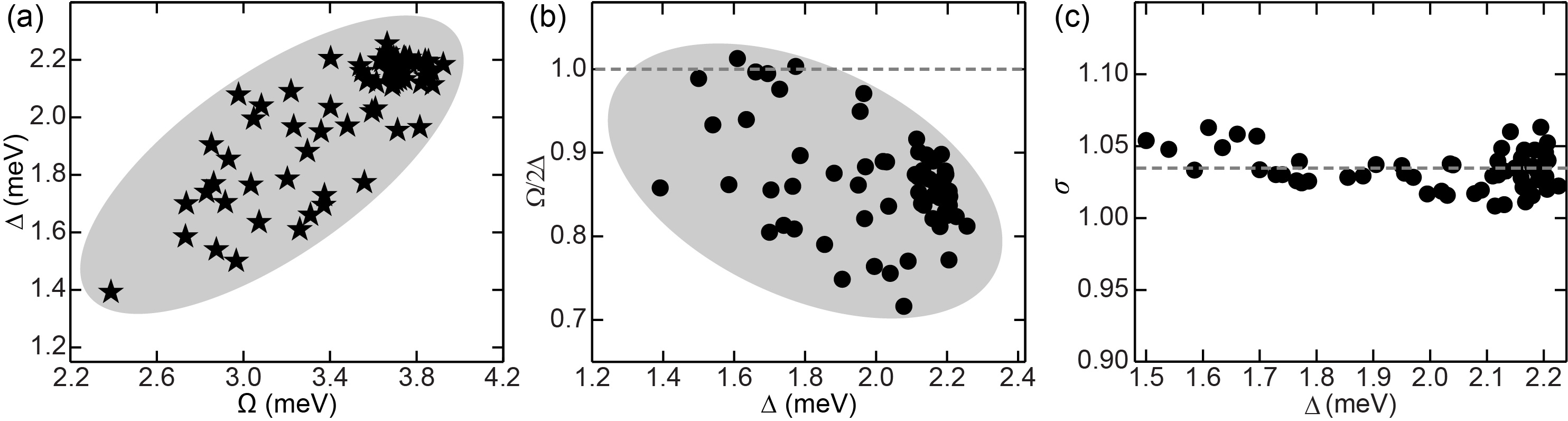}
  \caption{(a) Dependence of the superconducting gap magnitude $\Delta$ on the bosonic mode energy $\Omega$, acquired at 4.5 K. (b) $\Omega/2\Delta$ plotted as a function of $\Delta$. Note that $\Omega/2\Delta$ generally remains below 1 (gray dashed line) for the collected data. (c) Local normalized conductance ratio $\sigma$ vs.\ local $\Delta$. The dashed gray line is a guide to eye.
}
\end{figure*}

Having established the collective mode energy and its close connection to superconductivity, we now discuss the nature of this excitation. One candidate is the above-mentioned spin fluctuations, which have been found to link with the high-$T_{c}$ phase in pressurized FeSe crystals \cite{imai2009does}. Angle-resolved photoemission spectroscopy also revealed evidence of a short-ranged spin density wave in FeSe/SrTiO$_3$ films \cite{tan2013interface}. Furthermore, we here estimate $\Omega/k_{\mathrm{B}}T_{c}=4.7\pm0.2$ in FeSe, close to the ratios measured in other Fe-SCs by either electron tunneling or inelastic neutron scattering \cite{lumsden2010magnetism,qiu2009spin,wang2012close,song2013pairing}. These findings indicate that the observed mode probably corresponds to spin excitations. Based on the Eliashberg theory of superconductors with sign-reversing gaps, the energy of such a spin excitation must remain below the pairing-breaking energy, namely $\Omega/2\Delta<1$ \cite{eschrig2006effect}. Figure 4(b) demonstrates that the ratio $\Omega/2\Delta$ is generally below 1 in our data.

Alternatively, the mode may originate from phonons, although density functional theory reported insufficient electron-phonon coupling $\lambda$ ($\sim$ 0.17) to explain $T_{c}$ in FeSe \cite{subedi2008density}. Actually, such calculations have been recently questioned to significantly underestimate $\lambda$ in most correlated materials \cite{Yin2013correlation}. Additionally, the measurement of an iron isotope effect in FeSe highlights the importance of lattice vibrations in electron pairing \cite{Khasanov2010iron}. Finally, the bosonic mode energy $\Omega$ is reduced with local strain in Fig.\ 3(c). This is consistent with the identification of the observed bosonic mode as a phonon, because strain can modify the lattice vibrations and thus the phonon frequency. Further studies are needed to fully clarify the identity of the observed collective mode.

Finally we comment on the dependence of the local superconducting gap $\Delta$ on the bosonic mode energy $\Omega$. As has been demonstrated previously, $\Omega$ is locally anticorrelated with $\Delta$ in cuprate and iron pnictide superconductors \cite{niestemski2007distinct, jenkins2009imaging, fasano2010local, shan2012evidence, lee2006interplay}, which remains a counterintuitive puzzle and differs substantially from the present study. Balatsky \textit{et al}. tentatively addressed the puzzle in cuprates, proposing a local strong coupling model \cite{balatsky2006local}, where $\Delta(r)$ not only scales linearly with the boson mode energy $\Omega(r)$, but also correlates exponentially with the effective electron-boson coupling constant $g_{\mathrm{eff}}(r)$ as
\begin{equation}
\Delta(r)=\Omega(r)\mathrm{exp}\left(\frac{-1}{N_{\mathrm{0}}g_{\mathrm{eff}}(r)}\right)\,,
\end{equation}
where $N_{\mathrm{0}}$ is the density of states at the Fermi level. In addition, it was found that $g_{\mathrm{eff}}(r)$ is inversely proportional to $\Omega(r)$ by
\begin{equation}
g_{\mathrm{eff}}(r)=\frac{2g^{2}(r)}{\Omega(r)}\,.
\end{equation}
Here $g(r)$ is the local electron-boson coupling constant, which was found to be uncorrelated with $\Delta(r)$ in cuprates \cite{pasupathy2008electronic}. In order to examine $g(r)$ in FeSe, we quantified the normalized conductance ratio in Fig.\ 3(c)
\begin{equation}
\sigma=\frac{dI/dV(E_{\mathrm{hump}})}{dI/dV(E_{\mathrm{dip}})}\,,
\end{equation}
where a larger $\sigma$ means a stronger electron-boson coupling, i.e. larger $g(r)$. Figure 4(c) shows the measured $\sigma$ as a function of $\Delta$. No correlation is seen between the electron-boson coupling strength $\sigma$ and $\Delta$ within the experimental error. Therefore, we can approximately write $g(r)\equiv$ $g$ and arrive at the local gap magnitude $\Delta(r)$ as
\begin{equation}
\Delta(r)=\Omega(r)\mathrm{exp}\left(\frac{-\Omega(r)}{2N_{\mathrm{0}}g^{2}}\right)\,.
\end{equation}
Here $\Delta(r)$ shows a non-monotonic dependence on $\Omega(r)$. When $\Omega(r)$ is comparatively large (higher $T_{c}$), the exponential terms dominates, so the local enhancement in $\Omega(r)$ leads to weaker $g_{\mathrm{eff}}(r)$, and consequently to smaller $\Delta(r)$. This agrees excellently with the results in high-$T_{c}$ cuprates and iron pnictides \cite{niestemski2007distinct, jenkins2009imaging, fasano2010local, shan2012evidence, lee2006interplay}. However, when $\Omega(r)$ is smaller (lower $T_{c}$), as in the case of FeSe, the prefactor dominates and leads to a positive correlation between local $\Delta(r)$ and $\Omega(r)$, in agreement with our experimental data shown in Fig.\ 4(a).

In conclusion, our detailed STM/STS study has revealed a clear bosonic mode in FeSe. This mode is neither detectable above $T_{c}$ nor within the vortex cores. In strained regions, the local mode energy $\Omega$ shrinks as the superconducting gap $\Delta$ is also reduced, in contrast to other higher-$T_{c}$ superconductors where $\Omega$ anti-correlates with $\Delta$ \cite{niestemski2007distinct, jenkins2009imaging, fasano2010local, shan2012evidence, lee2006interplay}. A local strong coupling model \cite{balatsky2006local} explains the contrast well, and represents a unified theory of local electron-boson coupling in superconducting cuprates, iron pnictides, and iron selenides.

\begin{acknowledgments}
We thank J. Bauer and A. V. Balatsky for helpful discussions. This work was supported by National Science Foundation and Ministry of Science and Technology of China. C. L. S was partly supported by the Lawrence Golub Fellowship in Harvard University, and J. E. H. acknowledges support from the US National Science Foundation under Grant No. DMR-0847433. All STM topographic images were processed by WSxM software (www.nanotec.es).
\end{acknowledgments}
% Create the reference section using BibTeX:
%\bibliography{FeSeboson}
%

\end{document}